\documentclass[12pt]{article}
\usepackage{amsmath,amssymb,graphicx,slashed}
\usepackage{placeins}
\usepackage[tableposition=below]{caption}
\usepackage{longtable}
\usepackage{color}			
\usepackage{array}
\usepackage{verbatim}
\usepackage{braket}
\usepackage{bm}
\usepackage{breqn}
\usepackage{amsmath}
\usepackage{cite}
\usepackage{breqn}
\usepackage[utf8]{inputenc}
\usepackage{parselines}
\usepackage{float}
\usepackage{graphicx}
\usepackage{pgfplots}
\usepackage{subfig}
\usepackage{calrsfs}
\usepackage{slashed}
\usepackage{amsfonts}
\usepackage[cal=rsfso,calscaled=.96]{mathalfa}
\usepackage{subfig}
\usepackage[mathscr]{eucal}
\usepackage{sidecap}
\usepackage{wasysym}
\usepackage{caption}

\usepackage[
	colorlinks=true,
	citecolor=black,
	linkcolor=black,
	urlcolor=blue,
	hypertexnames=false]{hyperref}

\numberwithin{equation}{section}

\newcommand{\beq}{\begin{equation}}
\newcommand{\eeq}{\end{equation}}
\definecolor{MyRed}{rgb}{0.9,0.12,0.1}
\definecolor{MyBlue}{rgb}{0.1,0.12,0.9}

\newcolumntype{L}{>{$}l<{$}}

\begin{document}
\begin{titlepage}

\begin{center}

	{
		\LARGE \bf 
	Phenomenology of Bulk Scalar Singlets in the Randall Sundrum Model
		
	}
	
\end{center}
	\vskip .3cm
	
	\renewcommand*{\thefootnote}{\fnsymbol{footnote}}

\begin{center}
		
		\bf
		Fayez Abu-Ajamieh\footnote{\tt \scriptsize
		 \href{mailto:abuajamieh@ucdavis.edu}{abuajamieh@ucdavis.edu},
		 $^\dag$\href{houtz@ms.physics.ucdavis.edu}{houtz@ms.physics.ucdavis.edu},
		  $^\S$\href{zheng@ms.physics.ucdavis.edu}{zheng@ms.physics.ucdavis.edu}
		 },
		Rachel Houtz$^{\dag}$,
		%
	
		and Rui Zheng$^{\S}$
\end{center}
	{\it Department of Physics, University of California, One Shields Ave., Davis, CA 95616}

	\renewcommand{\thefootnote}{\arabic{footnote}}
	\setcounter{footnote}{0}


\begin{center} 

	{\it}

\end{center}


\centerline{\large\bf Abstract} 
\begin{quote}
We present a Randall-Sundrum toy model with an added scalar singlet that couples only to KK fermions in the bulk. Such a scalar would nontrivially affect radion phenomenology. In addition, we examine the radion phenomenology in light of the new scalar and show how this scalar could present another probe to search for the radion. 
\end{quote}

\end{titlepage}


\section{Introduction} 

The Randall-Sundrum (RS) model  \cite{Randall:1999ee} provides a simple solution for the hierarchy problem, explaining the disparity between the Electroweak (EW) scale $\sim 100$ GeV, and the Planck scale $\sim 10^{19}$ GeV. The LHC has generated renewed interest in RS models, and searches for evidence of extra dimensions are ongoing.

The recently-ruled out hints of a resonance at 750 GeV in the diphoton channel \cite{ATLAS:2015dp, CMS:2015dxe, CMS:2016owr} stirred interest in interpretations of the signal within extra dimensions scenarios. Reference \cite{Cai:2015hzc} presented an interpretation of the resonance as a scalar singlet that develops a VEV in one flat extra dimension. Other interpretations of the new scalar within the RS model were presented. In \cite{Csaki:2016kqr} the new particle was assumed to be a scalar that resides on another brane at $z  = z_{0}R'$ with $z_{0} < 1$. Interpretations of the new resonance being the radion or being radion-dominated were presented in \cite{Ahmed:2015uqt, Cox:2015ckc, Boos:2016ytd, Mahanta, Chaichian}, whereas \cite{Bauer:2016lbe} provided a solution through the introduction of a bulk scalar. Other proposals interpreting the resonance as a spin 2 graviton in warped geometries where presented in \cite{Arun, Falkowski, Hewett}.

Although the signal vanished in the subsequent analyses \cite{CMS2016}, the fact remains that new physics might first present itself in unexpected channels. Similar anomalous signatures could be the first hints of possible extra dimensions.

Here we present an RS model where we add a scalar singlet that couples only to vector-like or Kaluza-Klein (KK) fermions in the bulk. A bulk scalar that only couples to KK fermions means that all SM tree-level decays are forbidden, and only loop-level production and decay processes can occur. In this scenario, the bulk scalar can be produced by gluon fusion with KK quarks in the loop, and can only decay through a triangle diagram with KK quarks and leptons running in the loop to $WW$, $ZZ$, $gg$, $\gamma\gamma$ or $Z\gamma$. 

An important aspect of this model is that a bulk scalar can couple to the radion, which could lead to interesting radion phenomenology at the LHC. Scalar-radion associated production is possible through a box diagram or via gluon fusion followed by a tree-level decay of the off-shell scalar. In addition, if the radion is light enough, then the scalar could decay to a radion pair. If such a decay is kinematically allowed, it becomes significant and even dominant for moderate values of the radion VEV. Under these conditions, the scalar could present a new probe for radion searches.

This paper is organized as follows: In Section 2 we present our model and discuss the fermion and bulk scalar sectors, in Section 3 we investigate both the SM phenomenology of the model and that of the radion, including the constraints from LHC searches in the relevant channels, constraints from Electroweak Precision Observables (EWPO) and from KK graviton searches in Section 4 we discuss the radion discovery prospects at the LHC, and we discuss our results in Section 5.

\section{Model}
\subsection{Fermion Sector}

We consider a Randall-Sundrum model with the conformally flat metric:

\begin{equation} \label{metric}
ds^{2}=\Big(\frac{R}{z}\Big)^{2}(\eta_{\mu\nu}dx^{\mu}dx^{\nu}-dz^{2})
\end{equation}
This spacetime represents a slice of $AdS_{5}$, with boundaries between $R \sim 1/M_{Pl}$ and $R' \sim 1/\text{TeV}$. We place the fermion sector in the bulk and assume the Higgs is localized on the IR brane. We adopt the model in \cite{Csaki:2003sh} for the fermion sector and add a bulk scalar singlet $\Phi$ with a Yukawa term:

\begin{equation} \label{model}
S = \int d^{5}x\sqrt{g} \Big(\frac{i}{2}(\overline{\Psi}e^{M}_{a}\gamma^{a}D_{M}\Psi - D_{M}\overline{\Psi}e^{M}_{a}\gamma^{a}\Psi) -M_{KK} \overline{\Psi} \Psi - y_{f}\Phi\overline{\Psi}\Psi \Big)
\end{equation}
where $D_M=\partial_{M}+\frac12\omega^{ab}_M\sigma_{ab}$, and the $M$-index runs over the five-dimensional spacetime coordinates, $w^{ab}_M$ are the spin connections, and $M_{KK}$ is the fermion bulk mass. This action is explicitly written as:
\begin{equation} \label{explicit_model}
S=\int d^{5}x \Big( \frac{R}{z} \Big)^{4} \Big( -i\psi\sigma^{\mu}\partial_{\mu}\bar{\psi}\hspace{0.5 mm}-\hspace{0.5 mm} i\bar{\chi}\bar{\sigma}^{\mu}\partial_{\mu}\chi \hspace{0.5 mm} + \hspace{0.5 mm} \frac{1}{2}(\psi\overset{\leftrightarrow}\partial_{z}\chi \hspace{0.5 mm} - \hspace{0.5 mm} \bar{\chi}\overset{\leftrightarrow}\partial_{z}\bar{\psi}) \hspace{0.5 mm}+ \hspace{0.5 mm} (\frac{c\hspace{0.3 mm} + \hspace{0.3 mm} y_{f}R}{z})(\psi\chi \hspace{0.5 mm}+\hspace{0.5 mm} \bar{\chi}\bar{\psi}) \Big)
\end{equation}
where we used $\Psi=\begin{pmatrix} \chi \\ \bar{\psi}\end{pmatrix}$ and $c=MR$. This gives the following equations of motion:
\begin{equation} \label{EOM1}
-i\sigma^{\mu}\partial_{\mu}\bar{\psi}+\partial_{z}\chi+\frac{c-2}{z}\chi=0
\end{equation} 
\begin{equation} \label{EOM2}
-i\bar{\sigma^{\mu}}\partial_{\mu}\chi-\partial_{z}\bar{\psi}+\frac{c+2}{z}\bar{\psi}=0
\end{equation}
Using the KK decomposition
\begin{equation}\label{kk_cecomposition1}
\chi(x,z)=\sum_{n}g_{n}(z)\chi_{n}(x)
\end{equation}
\begin{equation} \label{kk_cecomposition2}
\bar{\psi}(x,z)=\sum_{n}f_{n}(z)\bar{\psi}_{n}(x)
\end{equation}
and the fact that the 4D spinors $\chi_{n}$ and $\psi_{n}$ obey the 4D Dirac equation, the equations of motion can be decoupled:
\begin{equation} \label{coupled_eq1}
f_{n}''-\frac{4}{z}f_{n}'+\Big( m_{n}^{2}-\frac{(c^{2}-c-6)}{z}\Big)f_{n}=0
\end{equation}
\begin{equation} \label{coupled_eq2}
g_{n}''-\frac{4}{z}g_{n}'+\Big( m_{n}^{2}-\frac{(c^{2}+c-6)}{z}\Big)g_{n}=0
\end{equation}
where $m_{n}$ are the masses of the KK modes. Before trying to solve the equations of motion, we need to address the boundary conditions. The possible boundary conditions are obtained by requiring that the fields vanish at the boundaries:
\begin{equation} \label{BC1}
\chi|_{R,R'} = 0
\end{equation}
\begin{equation} \label{BC2}
\bar{\psi}|_{R,R'} = 0
\end{equation}

Since the equations of motion must be satisfied at the boundaries and in the bulk, we can choose to impose Dirichlet's boundary condition on one of the fields, and extract the boundary conditions on the other field from the equations of motion. The set of consistent boundary conditions are:
\begin{equation} \label{consistent_BC1}
[\chi]_{R,R'}=0 \hspace{10 mm} \Rightarrow \hspace{10 mm} \Big[ \partial_{z}\bar{\psi}-\frac{c+2}{z}\bar{\psi} \Big]_{R,R'}=0
\end{equation}  
\begin{equation} \label{consistent_BC2}
[\bar{\psi}]_{R,R'}=0 \hspace{10 mm} \Rightarrow \hspace{10 mm} \Big[ \partial_{z}\chi+\frac{c-2}{z}\chi \Big]_{R,R'}=0
\end{equation}

Having obtained the boundary conditions, we can turn our attention to solving the equations of motion. The zero modes are the usual SM fermions. Their masses are assumed to be generated by the usual Higgs mechanism and not from the bulk, therefore $m_{0}$ in (\ref{coupled_eq1}) and (\ref{coupled_eq2}) is equal to zero. For the zero modes the equations decouple, and their solutions are given by:
\begin{equation} \label{g0_solution}
g_{0}(z) = A_{0}\Big(\frac{z}{R}\Big)^{2-c}
\end{equation}
\begin{equation} \label{f0_solution}
f_{0}(z) = C_{0}\Big(\frac{z}{R}\Big)^{2+c}
\end{equation}
The boundary conditions to be imposed on (\ref{g0_solution}) and (\ref{f0_solution}) should guarantee the zero modes are chiral, as required by SM fermions. We choose to impose the boundary conditions (\ref{consistent_BC1})\footnote{Note that the actual conditions to be imposed are: $\psi_{L}|_{R,R'}=\chi_{R}|_{R,R'} =0$ since the bulk gauge group is $SU(2)_{L} \times SU(2)_{R} \times U(1)$. For simplicity, in this paper we assume the LH and RH fermions are degenerate in the bulk, so that $\Psi_{L}=\Psi_{R}$.}. This gives the following normalized wavefunctions:
\begin{equation}   \label{f0_normalized_solution}
f_{0}(z)=0
\end{equation} 
\begin{equation}   \label{g0_normalized_solution}
g_{0}(z)=\frac{\sqrt{1-2c}}{R^{c}\sqrt{(R')^{1-2c}-(R)^{1-2c}}}\Big( \frac{z}{R}\Big)^{2-c}
\end{equation}
 Notice that the vanishing of $f_{0}$ does not imply the vanishing of $f_{n}$. These modes are coupled to $g_{n}$ through $m_{n}$, which are non-vanishing for the non-zero KK modes. Similarly, the solutions of the KK modes are given by:
\begin{equation}   \label{gn_normalized_solution}
g_{n}(z)=z^{5/2}A_{n}\Big(Y_{c-\frac{1}{2}}(m_{n}R)J_{c+\frac{1}{2}}(m_{n}z)-J_{c-\frac{1}{2}}(m_{n}R)Y_{c+\frac{1}{2}}(m_{n}z) \Big)
\end{equation}
\begin{equation}   \label{fn_normalized_solution}
f_{n}(z)=z^{5/2}A_{n}\Big(Y_{c-\frac{1}{2}}(m_{n}R)J_{c-\frac{1}{2}}(m_{n}z)-J_{c-\frac{1}{2}}(m_{n}R)Y_{c-\frac{1}{2}}(m_{n}z) \Big)
\end{equation}
where $A_{n}$ are overall normalization constants. In order to obtain the masses of the fermion KK modes, we simply impose the remaining boundary condition.

The parameter $c$ determines the localization of the zero-mode fermions. For $c_{L}>1/2$, the zero-modes are localized towards the Planck brane, whereas for $c_{L}<1/2$, the zero modes are localized near the TeV brane. Conversely, $c_{R}<-1/2 (>-1/2)$ implies that the right-handed fermions are localized near the Planck(TeV) brane. The CFT interpretation of this is that for $c_{L}>1/2$ and $c_{R}<-1/2$ fermions are elementary, whereas for $c_{L}<1/2$ and $c_{R}>-1/2$ they are composite.

We assume $c$ is equal for all KK fermions of the same handedness for simplicity. Furthermore, we shall assume $|c_{L}|$ = $|c_{R}|$, which means the left-handed (LH)
and right-handed (RH) KK masses are degenerate. The assumption of mass
degeneracy of the LH and RH fermions simply extends the case of SM
fermions (for which the masses of the LH and RH fermions is the same) to the
bulk. In addition, the assumption that all bulk fermions have the same value of $c$ is valid because the SM fermions are generated from the Higgs mechanism. The SM fermions are not zero modes of the bulk fermion wave functions, and therefore there is no need to match the bulk fermion wavefunctions with the SM masses. 
\subsection{Scalar Sector}
The action of the bulk scalar is given by:
\begin{equation} \label{scalar_sector}
S_{\Phi}=\int d^{5}x \sqrt{g} \Big( \frac{1}{2}g^{MN}\partial_{M} \Phi\partial_{N} \Phi - \mu^{2} \Phi^{2})  \mp \frac{1}{2} \int d^{4}x \sqrt{g_{0,1}}M_{0,1}^{2} \Phi^{2}
\end{equation}
where $g_{0,1}$ are the induced metrics on the UV and IR branes respectively, and $M_{0,1}$ are the brane-localized mass terms. Brane-localized masses control the size of the extra dimension by tuning the brane tension. They are introduced in order to avoid constraining the size of the extra dimension. One might worry that introducing the brane mass terms leads to fine-tuning. In fact no such fine-tuning is necessary in order to obtain a proper size of the extra dimension, as $R'$ is not sensitive to $M_{0,1}$. Throughout this paper, we fix $M_{0,1}^{2}$ to 1 and 4 in units of $R = 1$.

The zero mode of $\Phi$ is of most interest to this analysis. The solutions of the equation of motion of the scalar action is given by:

\begin{equation} \label{scalar_solution}
\varphi_{n}(z)=z^{2} \Big(  C_{n} J_{\beta}(m_{\phi}^{(n)}) + D_{n} Y_{\beta}(m_{\phi}^{(n)})  \Big)
\end{equation}
where $\beta= \sqrt{4+2 \mu^{2}R^{2}}$. Notice that $\beta > 2$. The boundary conditions are given by:

\begin{equation} \label{scalar_BC1}
[\varphi_{n}'(z) - M_{UV}^{2} \varphi_{n}(z)]_{R}=0
\end{equation}

\begin{equation} \label{scalar_BC2}
[\varphi_{n}'(z) - \Big(\frac{R}{R'} \Big)M_{IR}^{2} \varphi_{n}(z)]_{R'}=0
\end{equation}
The KK masses can be obtained by imposing the boundary conditions. Note that the $m_{\phi_{0}}$ is determined by $R'$ and $\beta$, so we choose two benchmark points and calculate the corresponding zero-mode mass. The benchmark points are summarize in Table \ref{table1}.

\begin{center}
\begin{tabular}{L | L | L} 
& \text{Point 1} & \text{Point 2} \\
\hline
\hline
& &\\
M_{0}^{2} (R =1) & 1 & 1/R \\
& &\\
M_{1}^{2} (R = 1) & 4 & 4/R \\
& &\\
R' (GeV^{-1})& 1/767 & 1/550 \\
& &\\
\beta & 2.1 & 2.5 \\
\hline
\hline
& &\\
m_{\phi_{0}} (\text{GeV}) & 600 & 1000

\end{tabular}
\captionof{table}{The two sets of parameters with the corresponding bulk scalar zero-mode mass.}\label{table1}
\end{center}

\subsection{Scalar Interaction}
In the 4D effective theory, the Yukawa term of the lowest scalar mode can be written explicitly as:

\begin{equation} \label{S_Yukawa}
S_{Yukawa}= -y_{\text{eff}} \int d^{4}x \phi_{0} (x) \Big( \sum_{n=1} \overline{\Psi}_{0}\Psi_{n}  +  \sum_{n=1} \overline{\Psi}_{n}\Psi_{n} +  \sum_{\substack{n,l=1\\ n \neq l}} \overline{\Psi}_{l}\Psi_{n}  \Big) 
\end{equation}
where we have assumed universal coupling for all fermions and all KK modes for simplicity. Notice the absence of any $\overline{\Psi}_{0}\Psi_{0}$, which means the scalar cannot be produced by or decay to SM fermions at tree level. Also, since the KK fermion masses $\sim x_{i}/R'$, where $x_{i}$ is the zero of the appropriate Bessel function, and the natural mass of $m_{\phi_{0}} \sim$ TeV, the masses of the first KK fermions are of $O($a few $\text{TeV})>m_{\phi_{0}}$. This means all tree-level scalar decays are kinematically forbidden.

Naively, (\ref{S_Yukawa}) implies $\phi_{0}$ cannot decay to a Higgs pair through triangle diagrams, since the KK fermions only couple to the scalar, while the Higgs is localized on the TeV brane and couples of SM fermions. Actually, as shown by \cite{Cai:2015hzc}, if the bulk scalar develops a VEV, then it can mix with the Higgs through the potential term:

\begin{equation} \label{Higgs_mixing}
V_{\text{mixing}} = \frac{\lambda_{\phi h}}{2}(v+h)^{2}(v_{\phi} + \phi)^{2}
\end{equation}
where $v_{\phi}$ is the scalar VEV. In order to avoid introducing too-large a correction to the Higgs sector, we shall take $\lambda_{\phi h} \ll 1$. 

An interesting aspect of this model is that the scalar can couple to the radion, which could provide a discovery channel for the radion at the LHC. The radion is expected to be the lightest particle in the RS model, and so it deserves extra attention. The radion field can be parametrized as perturbations about the background metric \cite{Csaki:2007ns}:
\begin{equation} \label{metric-perturbation}
ds^{2} = \Bigg( \frac{R}{z}\Bigg)^{2} \Big( e^{-2F(x,z)} \eta_{\mu\nu} dx^{\mu} dx^{\nu} - (1+2F(x,z))^{2} dz^{2}\Big)
\end{equation}
In the limit of a small backreaction, the normalized wavefunction of the radion is given by:

\begin{equation} \label{radion_solution}
F(x,z) = \frac{r(x)}{\Lambda_{r}} \Bigg( \frac{z}{R'} \Bigg)^{2}
\end{equation}
where $\Lambda_{r} \equiv \frac{\sqrt{6}}{R'}$ is the radion VEV and $r(x)$ is the normalized 4D radion field. To find the radion-scalar coupling, we use the metric (\ref{metric-perturbation}) in the scalar sector given by eq. (\ref{scalar_sector}). After performing the KK decomposition and integrating out the extra dimension, we obtain the radion-scalar effective coupling. The effective 4D radion scalar coupling is shown below:

\begin{SCfigure}[][h!]
  \begin{minipage}{.5\textwidth}
    \centering
    \includegraphics[scale=0.5]{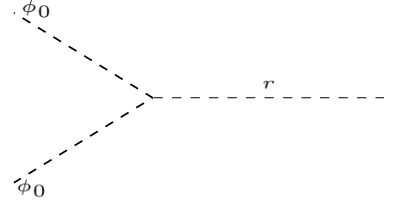}
  \end{minipage}%
  \begin{minipage}{.5\textwidth}
  \begin{equation} \label{SSr}
      \equiv \frac{1}{\Lambda_{r}} \Bigg( \frac{R}{R'} \Bigg)^{2} I(c,\beta)
    \end{equation}
  \end{minipage}
\end{SCfigure}
\noindent where $I(c,\beta)$ is a numerical constant coming from the fermion loop integration:

\begin{equation}\label{SSr Integral}
I(c,\beta) = \int_{R}^{R'} \Big(\frac{R}{z}\Big) \Big[ 3 \varphi_{n}^{'2} +\frac{\beta^{2} - 4}{z^{2}}\varphi_{n}^{2} \Big] 
\end{equation}

\section{Phenomenology}

The scalar $\phi_{0}$ can be produced by gluon fusion with KK quarks in the loop, and can decay through a triangle diagram with KK quarks and leptons in the loop. In both cases, SM fermions cannot run in the loop as can be seen from (\ref{S_Yukawa}). The field $\phi_{0}$ can have SM decays as well as non-SM decays. We treat each case separately. 

Before treating the phenomenology of this model, we need to estimate the number of KK fermions running in the loop. To do this, we invoke unitarity. In $4+d$ dimensions, Yang-Mills theories are non-renormalizable, and the gauge coupling has a mass dimension $-d/2$. Thus, such a theory should be treated as an effective theory up to some UV cutoff scale $\Lambda$. The effective theories can be used as long as the scattering amplitudes remain unitary. 

In model with higher dimensions, the large number of KK modes usually leads to unitarity violation. Keeping in mind that the effective theory description is valid up to some scale $\Lambda$, one can calculate the maximum number of KK modes to include before unitarity is violated. This was calculated in \cite{Chivukula:2003kq} for the case of a flat extra dimension. The authors showed that the unitarity of gluon scattering amplitude imposes the most stringent constraint and found $N_{KK} = 2$ for a single extra dimension. The RS case, on the other hand, has never been calculated, so we resort to Naive Dimensional Analysis (NDA) to estimate $N_{KK}$. Surprisingly, NDA gives the same result obtained for flat extra dimension, so we set $N_{KK} = 2$. A more detailed explanation of our estimation of $N_{KK}$ is in Appendix \ref{NKK}.
\subsection{SM Decays}

\begin{figure}
\centering
\includegraphics[scale = 0.3]{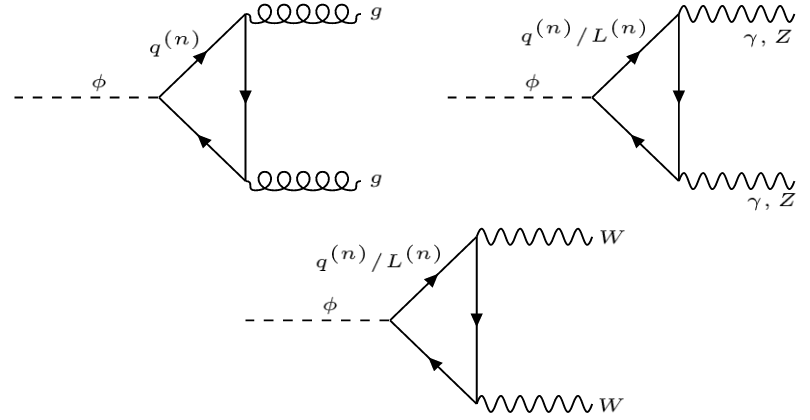}
\caption{Scalar SM decays to $gg$, $ZZ$, $WW$, $\gamma\gamma$ and $Z\gamma$}
\label{triangle_diagrams}
\end{figure}

The scalar $\phi_{0}$ can decay to $gg$, $ZZ$, $W^{+}W^{-}$, $\gamma \gamma$ or $Z\gamma$. The decays happen through triangle diagrams as shown in Fig.~\ref{triangle_diagrams}. 

The scalar $\Phi$ only couples to KK fermions, while the Higgs is assumed to couple only to SM fermions (the zero modes). This means $\phi_{0}$ cannot decay to $hh$, whether at tree level or through the triangle diagrams. The only way for $\phi_{0}$ to mix with the Higgs is through (\ref{Higgs_mixing}), which is assumed to be small. Another consequence of $\Phi$'s coupling exclusively to KK fermions is that $\Phi$ cannot decay to SM fermions at tree level or via gauge boson loops. 

From the form of (\ref{S_Yukawa}), $\phi_{0}$ cannot decay at tree level to a SM fermion and a KK fermion. This would not be kinematically allowed, alleviating constraints from top searches, Drell-Yan, or similar processes.

If the radion is light enough ($< m_{\phi_{0}}$), then $\phi_{0}$ can also decay to a radion pair through a similar triangle diagram\footnote{In general, the radion could couple to gluons through the trace anomaly term $\frac{r}{\Lambda}\beta \frac{\alpha_{s}}{8\pi}G_{\mu\nu}G^{\mu\nu}$ where $\beta$ is the beta fucntion of the gluon. Here, as can be seem from \ref{SSr}, the lack of a $\phi_{0}rr$ vertex means that radions cannot run in the loop in the processes in Fig.~\ref{triangle_diagrams}}. For now we shall assume the radion mass is heavy so that it decouples from the theory and focus on the scalar SM decays. Those decays are similar to the well-studied SM Higgs decays. Using the notation of \cite{Altmannshofer:2015xfo, Carmi:2012in}:

\begin{dmath} \label{effective_lagrangian}
\mathcal{L} \supset \kappa_{g} \phi_{0} G_{\mu\nu}^{a}G^{a \mu \nu} + \kappa_{\gamma} \phi_{0} F_{\mu \nu} F^{\mu \nu} + \kappa_{Z}  \phi_{0} Z_{\mu \nu} Z^{\mu \nu} + \kappa_{Z \gamma} \phi_{0} Z_{\mu \nu}F^{\mu \nu} \\ + \kappa_{W} \phi_{0} W_{\mu\nu}^{+} W^{-\mu\nu}
\end{dmath}
The definitions of the effective couplings are given in Appendix \ref{effective_couplings}. In our calculation, we use the narrow width approximation:

\begin{equation} \label{narrow_width}
\sigma(pp \rightarrow \phi_{0} \rightarrow X_{1}X_{2}) = \sigma (pp \rightarrow \phi_{0}) \times Br (\phi_{0} \rightarrow X_{1}X_{2})
\end{equation}
The production cross section can be obtained from:

\begin{equation} \label{production_xsection}
 \sigma (pp \rightarrow \phi_{0}) = \Bigg( \frac{\kappa_g(\phi_{0})}{\kappa_g(h)} \Bigg)^{2} \times \sigma (gg \rightarrow h)
\end{equation}
where $\sigma (gg \rightarrow h)$ is the SM Higgs production cross section at $m_{h} = 600$ and 1000 GeV respectively. The values we use are $\sigma_{600}^{13\,\text{TeV}} = 1000.1$ fb and $\sigma_{1000}^{13\,\text{TeV}} = 184.5$ fb \cite{Heinemeyer:2013tqa, CERN0}. We can use (\ref{effective_lagrangian}) to calculate the decay widths and branching ratios. We find:
\begin{equation} \label{diphoton_decay}
\Gamma(\phi_{0} \rightarrow \gamma\gamma) = \frac{\kappa_{\gamma}^{2}m_{\phi}^{3}}{4 \pi}
\end{equation}

\begin{equation} \label{gluons_decay}
\Gamma(\phi_{0} \rightarrow gg) = \frac{2\kappa_{g}^{2}m_{\phi}^{3}}{ \pi}
\end{equation}

\begin{equation} \label{WW_decay}
\Gamma(\phi_{0} \rightarrow W^{+}W^{-})= \frac{\kappa_{W}^{2} m_{\phi}^{3}}{16 \pi}\sqrt{1-\frac{4m_{W}^{2}}{m_{\phi}^{2}}} \Bigg[ 2 \Big( 1-\frac{2m_{W}^{2}}{m_{\phi}^{2}}\Big)^{2} + \frac{4m_{W}^{4}}{m_{\phi}^{4}} \Bigg]
\end{equation}

\begin{equation} \label{ZZ_decay}
\Gamma(\phi_{0} \rightarrow ZZ) =\frac{\kappa_{Z}^{2} m_{\phi}^{3}}{8 \pi}\sqrt{1-\frac{4m_{Z}^{2}}{m_{\phi}^{2}}} \Bigg[ 2 \Big( 1-\frac{2m_{Z}^{2}}{m_{\phi}^{2}}\Big)^{2} + \frac{4m_{Z}^{4}}{m_{\phi}^{4}} \Bigg]
\end{equation}

\begin{equation} \label{photon_Z_decay}
\Gamma(\phi_{0} \rightarrow Z\gamma) = \frac{\kappa_{Z\gamma}^{2} m_{\phi}^{3}}{8 \pi} \Big( 1-\frac{m_{Z}^{2}}{m_{\phi}^{2}}\Big)^{3}
\end{equation}
The branching ratios are constant throughout the parameter space examined here. The branching ratios for both benchmark points are the same, and are given by:
\begin{equation} \label{gg_Br}
Br(\phi_{0} \rightarrow gg) \hspace{1 mm} \simeq \hspace{1 mm} 96.9 \hspace{1 mm} \%
\end{equation}

\begin{equation}  \label{WW_Br}
Br(\phi_{0} \rightarrow WW) \hspace{1 mm} \simeq \hspace{1 mm} 1.9 \hspace{1 mm} \%
\end{equation}

\begin{equation}  \label{ZZ_Br}
Br(\phi_{0} \rightarrow ZZ) \hspace{1 mm} \simeq \hspace{1 mm} 0.71 \hspace{1 mm} \%
\end{equation}

\begin{equation}  \label{diphoton_Br}
Br(\phi_{0} \rightarrow \gamma\gamma) \hspace{1 mm} \simeq \hspace{1 mm} 0.35 \hspace{1 mm} \%
\end{equation}

\begin{equation}  \label{photon_Z_Br}
Br(\phi_{0} \rightarrow Z\gamma) \hspace{1 mm} \simeq \hspace{1 mm} 0.09 \hspace{1 mm} \%
\end{equation}
As expected, $\phi_{0}$ decays predominantly to gluons. We now investigate the LHC bounds on the parameter space for the two benchmark points in Table \ref{table1} and leave the parameters $c$ and $y_{\text{eff}}$ free. The latest bounds at 95\% C.L. from the LHC 13 TeV\footnote{Some of the data is only available for $\sqrt{s} = 8$ TeV. So we scale them using the parton luminosity ratios} run for a $600$ $(1000)$ GeV scalar are given by  \cite{ATLAS1,ATLAS2,ATLAS3,CMS1,ATLAS4,ATLAS5,Barman}:

\begin{equation} \label{bound_ZZ}
\sigma (pp \rightarrow \phi_{0} \rightarrow ZZ) < 500  \hspace{1 mm} (20) \hspace{1 mm} \text{fb},
\end{equation}

\begin{equation}  \label{bound_WW}
\sigma (pp \rightarrow \phi_{0} \rightarrow W^{+}W^{-}) < 500  \hspace{1 mm} (50) \hspace{1 mm} \text{fb},
\end{equation}

\begin{equation}  \label{bound_Zphoton}
\sigma (pp \rightarrow \phi_{0} \rightarrow Z\gamma) < 40  \hspace{1 mm} (25) \hspace{1 mm} \text{fb},
\end{equation}

\begin{equation}  \label{bound_diphoton}
\sigma (pp \rightarrow \phi_{0} \rightarrow \gamma \gamma) < 5  \hspace{1 mm} (2) \hspace{1 mm} \text{fb},
\end{equation}

\begin{equation}  \label{bound_jj}
\sigma (pp \rightarrow \phi_{0} \rightarrow jj) < 12.4  \hspace{1 mm} (5.9) \hspace{1 mm} \text{pb},
\end{equation}

\begin{figure}
\centerline{%
\includegraphics[width=0.6\textwidth]{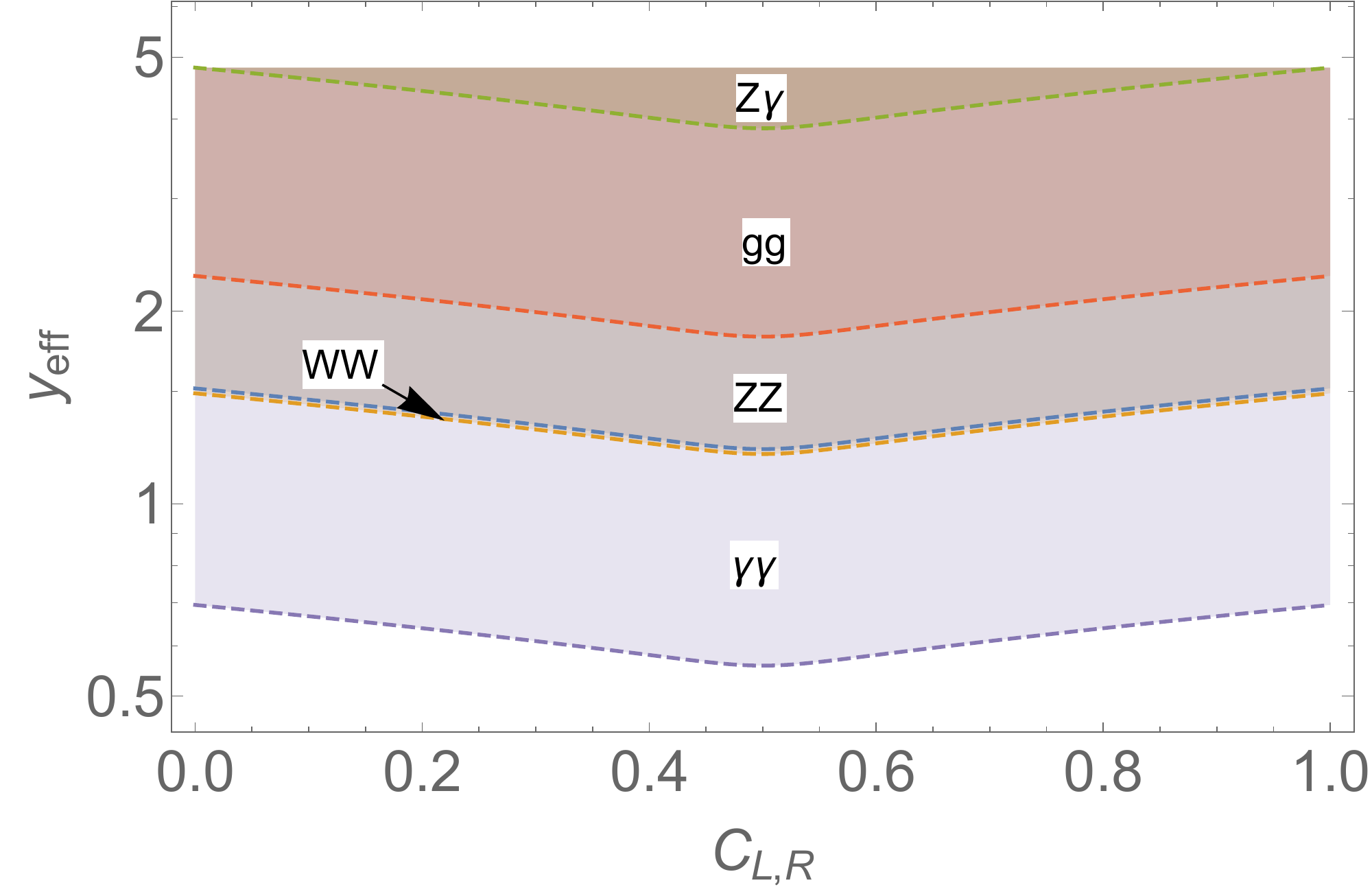}%
\includegraphics[width=0.6\textwidth]{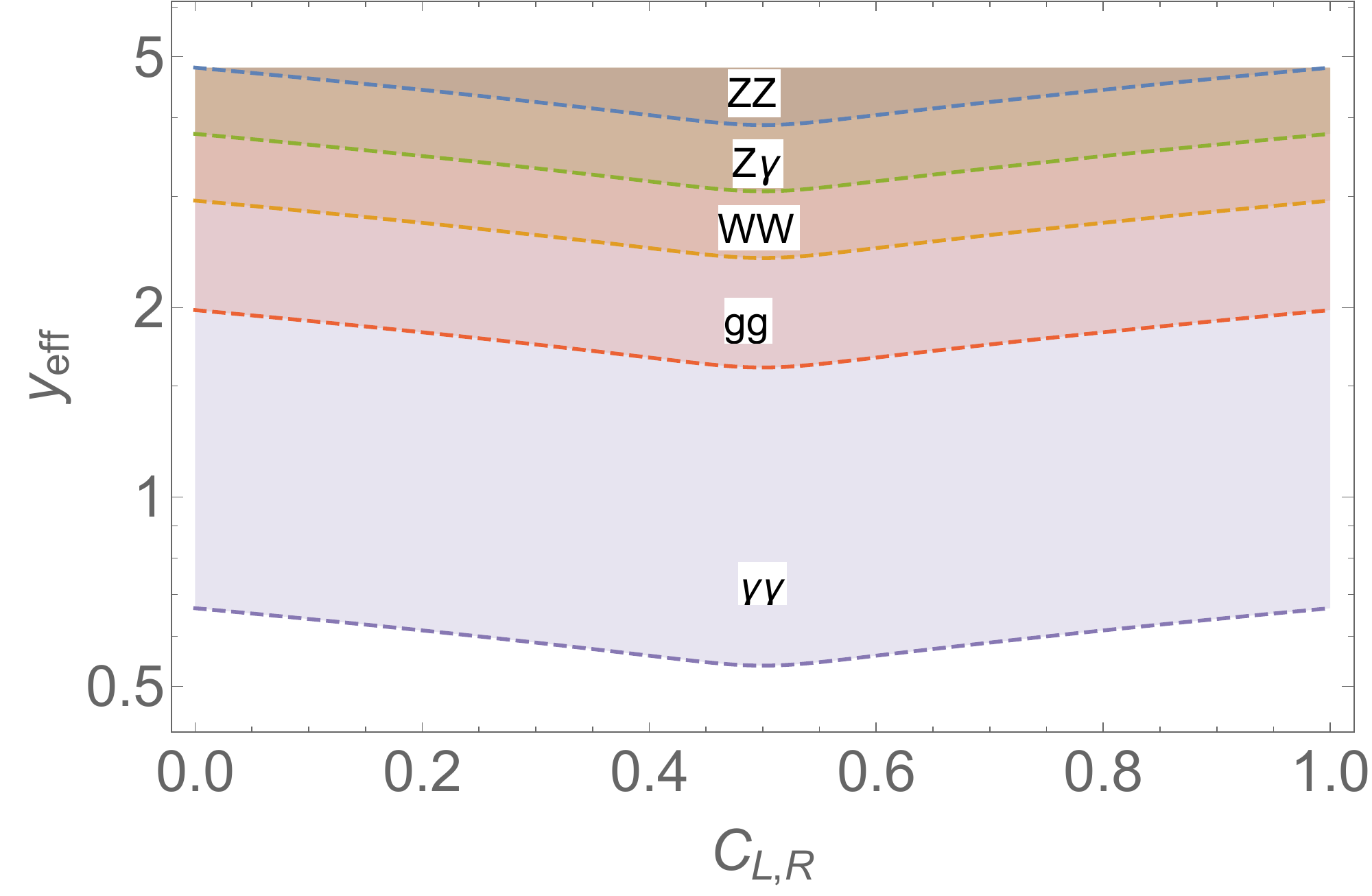}%
}%

\caption{Excluded regions plots in $y_{\text{eff}}- |c_{L,R}|$ parameter space for  benchmark points 1 (left) and 2 (right) on a log scale.}
\label{fig:exclusion1}
\end{figure}

We plot these bounds on the $y_{\text{eff}} - c$ parameter space. As can be seen from Fig.~\ref{fig:exclusion1}, the most stringent bounds comes from the diphoton channel. Coupling values $y_{\text{eff}} \lesssim 0.6$ are excluded for the overall range of $c$. As we will see shortly, however, these constraints are alleviated once the decay to radions in turned on.

\subsection{Radion Phenomenology}

We now assume the radion mass is less than $m_{\phi_{0}}/2$. In this case, $\phi_{0}$ can decay to a pair of radions through a triangle diagram similar to Fig.~\ref{triangle_diagrams}. This leads to potentially interesting radion phenomenology, since the branching ratio of the $\phi_{0}$ decay to radions becomes significant. In this calculation, we shall not adhere to any particular model for the radion, although we assume that the radion wavefunction has the general form in (\ref{radion_solution}). Using this framework, $\Lambda_{r}$ is now a free parameter.

\begin{figure}
\centerline{%
\includegraphics[width=0.6\textwidth]{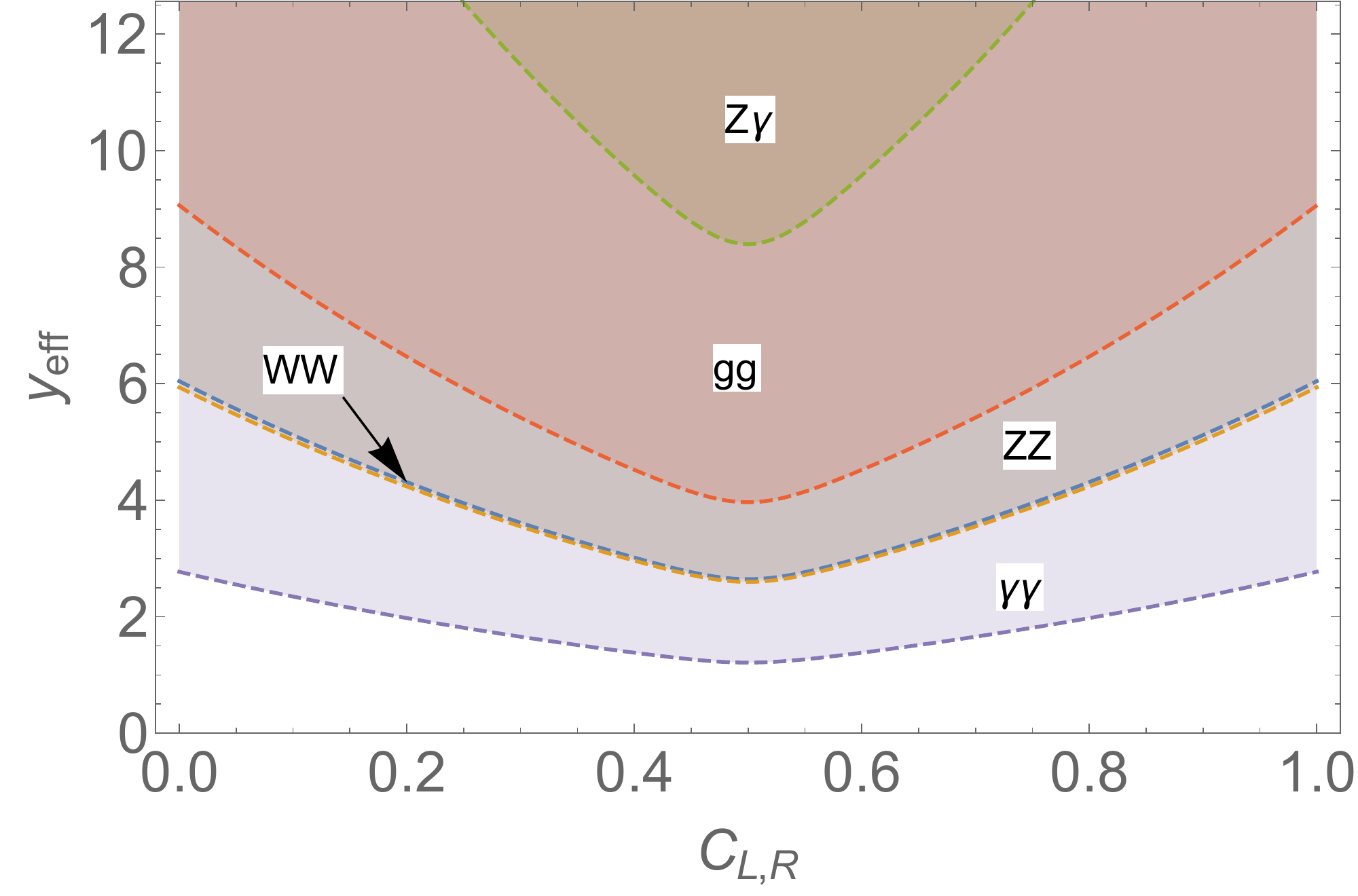}%
\includegraphics[width=0.6\textwidth]{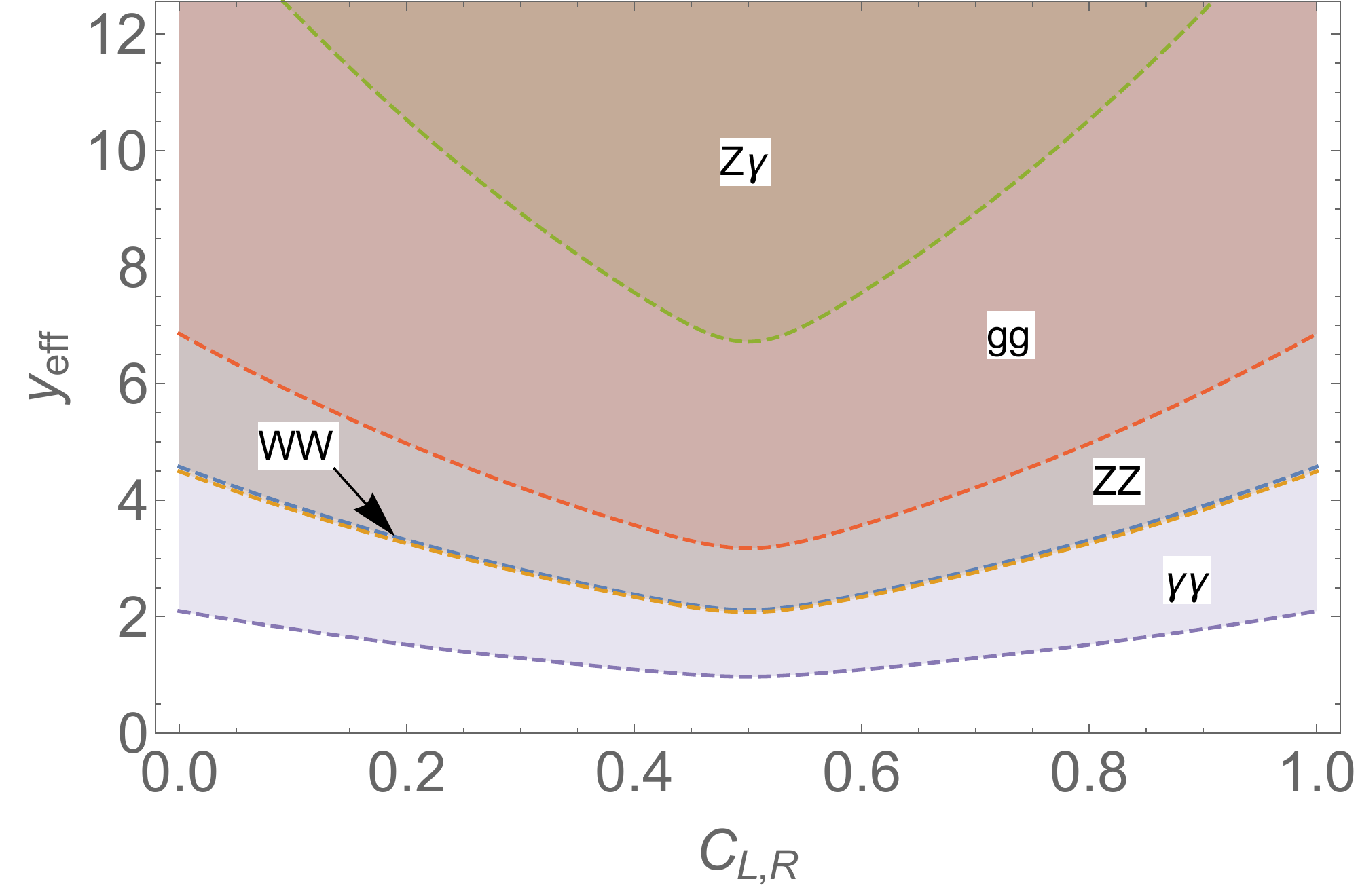}%
}%

\caption{Excluded regions plots in $y_{\text{eff}} - |c_{L,R}|$ parameter space for the parameters in benchmark point 2 with the decay to radions turned on. Here we have $\Lambda_{r} = 3$ TeV and $m_{r} = 300$ GeV (left) and $450$ GeV (right).}
\label{fig:exclusion2}
\end{figure}

First we re-examine the LHC constraints from the 13 TeV run after the radion channel is turned on. With the decay to radions turned on, we have two additional parameters, the radion mass $m_{r}$ and the radion VEV $\Lambda_{r}$. For simplicity, we choose benchmark point 2, fix $\Lambda_{r} = 3$ TeV and calculate the constraints for $m_{r} = 300$ and $450$ GeV respectively. The modified constraints are shown in Fig.~\ref{fig:exclusion2}.

\begin{figure}[H]
\centerline{%
\includegraphics[width=0.6\textwidth]{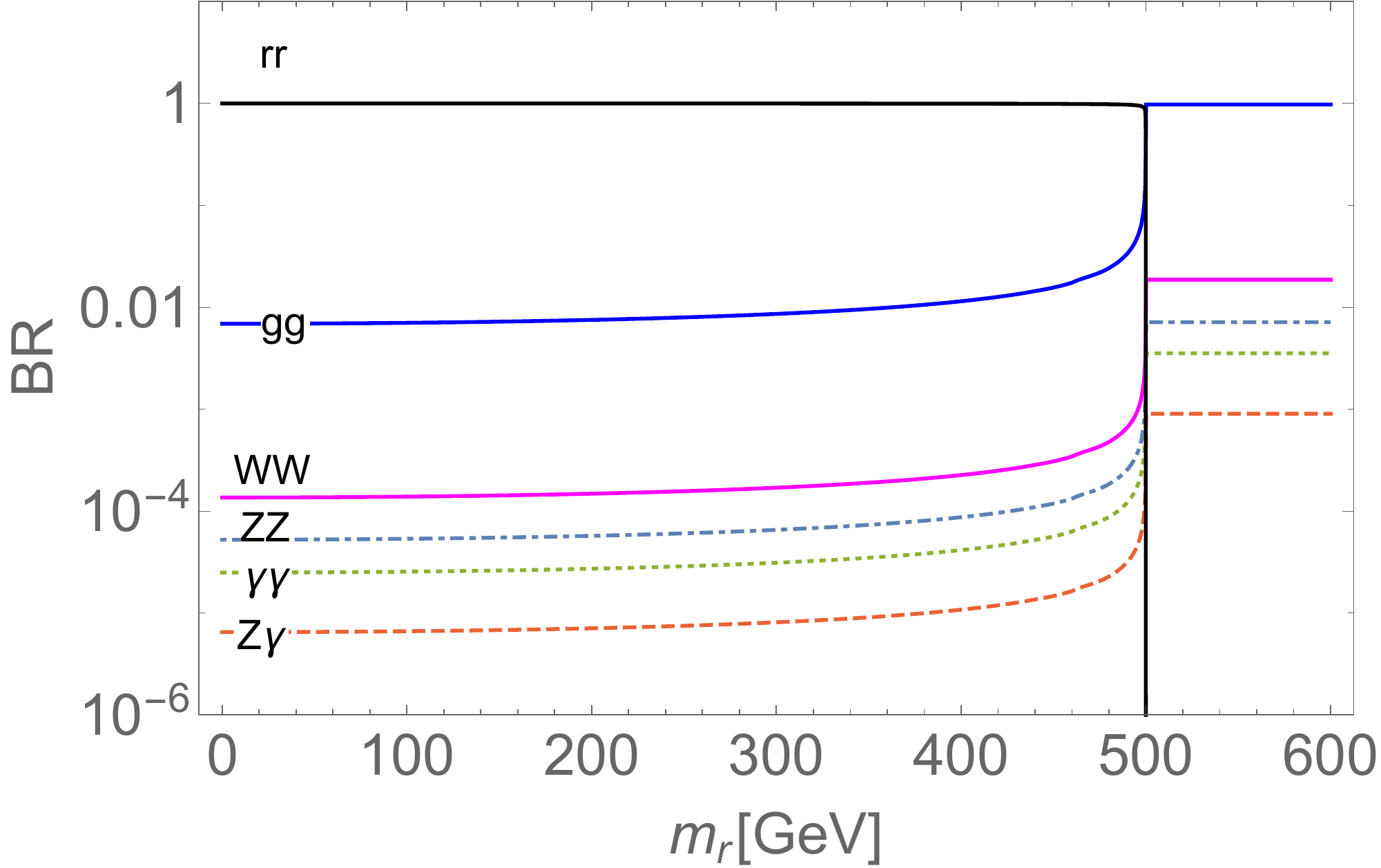}%
\includegraphics[width=0.6\textwidth] {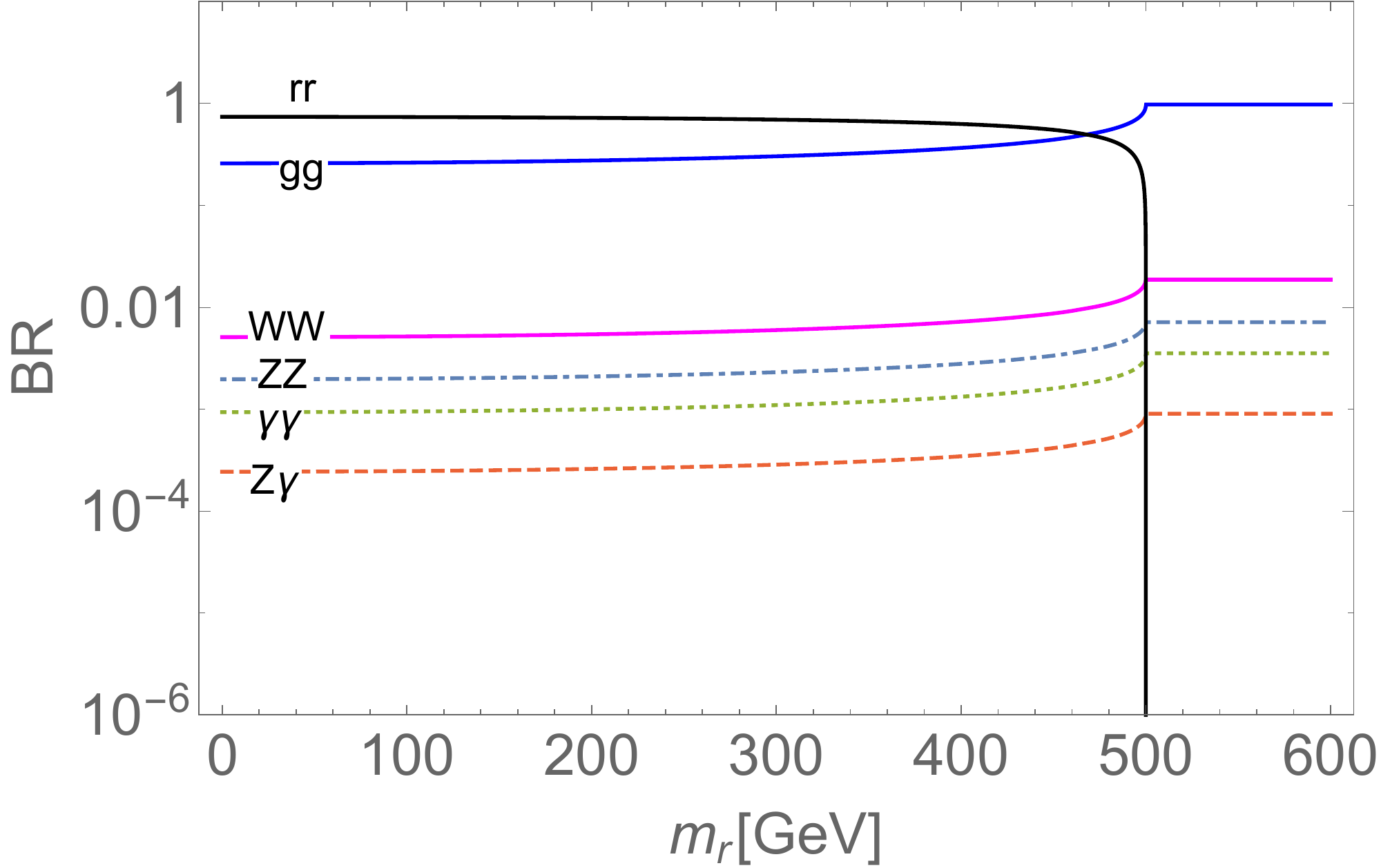}%
}%

\bigskip
\caption{Branching ratios of $\phi_{0}$ decay for benchmark point 2 corresponding to $m_{\phi_{0}} = 1000$ GeV with $c = 0.487$, for $3$ TeV (left) and $8$ TeV (right)}
\label{fig:scalar_BR}
\end{figure}

Once again the $\gamma\gamma$ channel imposes the most significant constraints. As expected, the constraints on $y_{\text{eff}}$ are relaxed since the radion channel is now competing with the other SM channels. We can also see that the bounds are more stringent for a radion mass of 450 GeV compared to 300 GeV. This is because the branching ratio becomes smaller for a heavier mass (\textit{c.f.} Fig.~\ref{fig:scalar_BR}), which means that the cross section of the other decays becomes larger and hence more constrained by the LHC bounds.

\begin{figure}
\includegraphics[scale = 0.5]{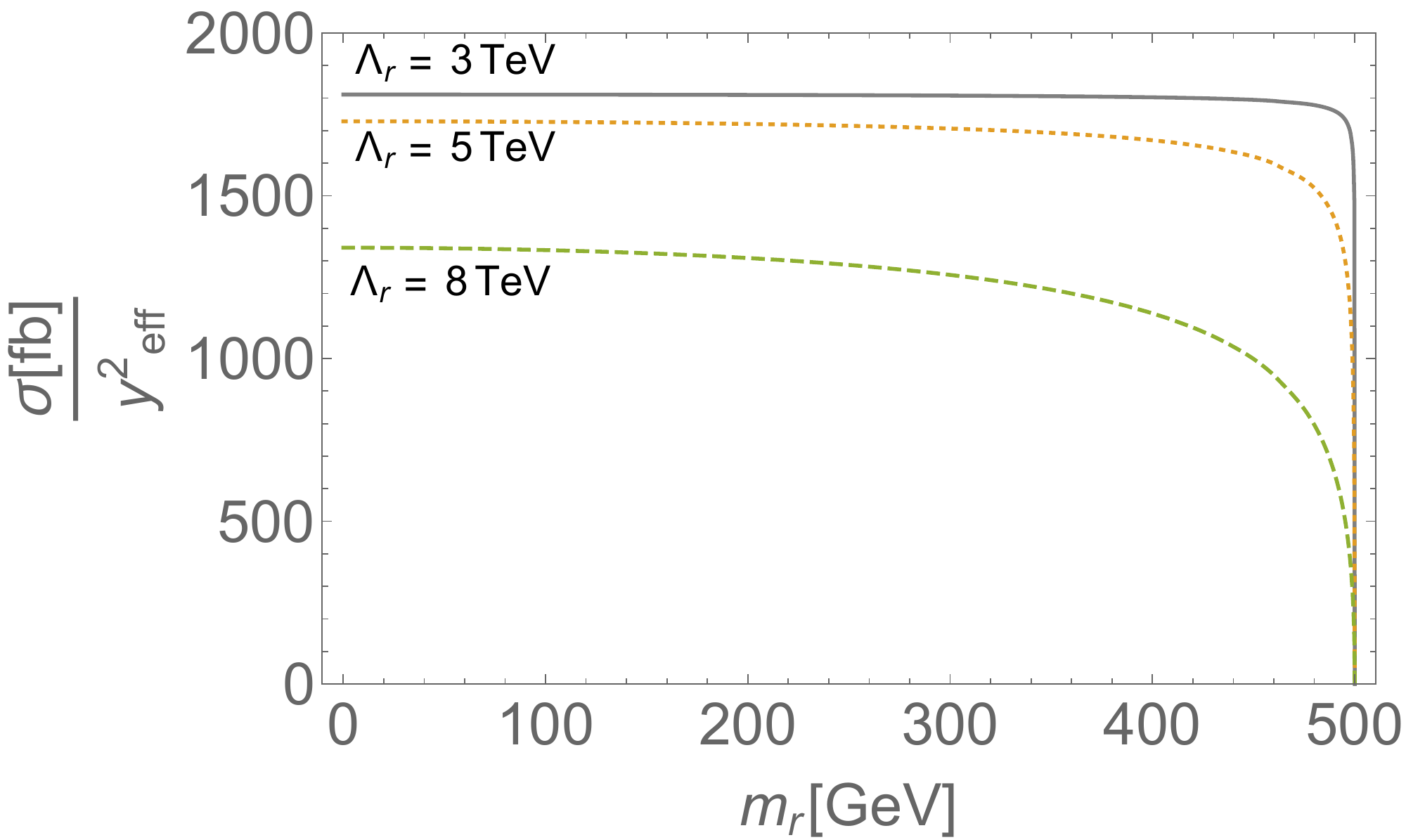}
\centering
\caption{Radion pair production cross section for $3,\hspace{1 mm} 5$ and $8 \hspace{1 mm}$ TeV. The cross section vanishes at 500 GeV, which corresponds to $m_{\phi_{0}}/2$.}
\label{fig:rr_production}
\end{figure}

We calculate the production cross section of the radion pair using the narrow width approximation (\ref{narrow_width}). Here we fix $m_{\phi_{0}} = 1000$ GeV and  calculated its decay width to radions explicitly using dimensional regularization in the $\overline{MS}$ scheme at a renormalization scale $ = m_{\phi_{0}}$. Fig.~\ref{fig:rr_production} shows the production cross section of the radion pair for $\Lambda_{r} =$ 3, 5 and 8 TeV. As can be seen from the plot, the production cross section changes slowly for $m_{r} \lesssim 450$ GeV and then drops quickly to zero at $m_{r} > 500$ GeV when the decay becomes kinematically forbidden. For a radion mass $\lesssim 450$ GeV the production cross section can be significant even for small  values of the coupling constant.

In Fig.~\ref{fig:scalar_BR} we show the branching ratios of $\phi_{0}$ decay. Here we pick benchmark point 2 and fix $c = 0.487$ (the choice of this value is explained in Section 3.3). We show the branching ratios as a function of the radion mass for $\Lambda_{r} = \sqrt{6}/R'$, 3, 5 and 8 TeV. As can be seen from the plots, the decay to a pair of radions is the dominant channel over most of the mass range. As $\Lambda_{r}$ becomes larger, however, this channel becomes more suppressed and the $gg$ channel starts to compete with it.

\subsection{Bounds from Electroweak Precision Observables and Graviton Searches}
Here we investigate the constraints Electroweak Precision Observables (EWPO) place on our model. We will focus on the oblique parameters, especially the $S$ parameters as they are the most problematic in the RS model. It is known that the original RS model with fermions localized on the TeV brane, leads to large negative contributions to the $S$ parameter \cite{Csaki:2002}. On the other hand, as noted in \cite{Agashe:2003}, localizing the fermions on the UV brane leads to positive contribution to the $S$ parameter. This means if the fermions are allowed to propagate in the bulk, there will be a region where the contribution to the $S$ parameter is vanishing. This argument was used in \cite{Cacciapaglia:2005} in order to solve the $S$ problem in the Higgsless model and remains valid in this model as well. According to their results, for $(R/R')^{2c-1} \ll 1$, the contribution to the $S$ parameter is given by:

\begin{equation} \label{SCont1}
S = \frac{6 \pi}{g^{2} \log{\frac{R'}{R}}} \Bigg(1-\frac{4}{3}\frac{2c-1}{3-2c} \Big(\frac{R}{R'} \Big)^{2c-1} \log{\Big(\frac{R}{R'}\Big)} \Bigg)
\end{equation}
while $T \approx U \approx 0$. On the other hand, for $c \approx 1/2$, the leading contribution to the $S$ parameter is given by:

\begin{equation}\label{SCont2}
S \approx \frac{2 \pi}{g^{2} \log{\frac{R'}{R}}} \Bigg(1+(2c-1) \log{\Big( \frac{R'}{R} \Big)}  \Bigg)
\end{equation}

\begin{figure}[H]
\centerline{%
\includegraphics[width=0.6\textwidth]{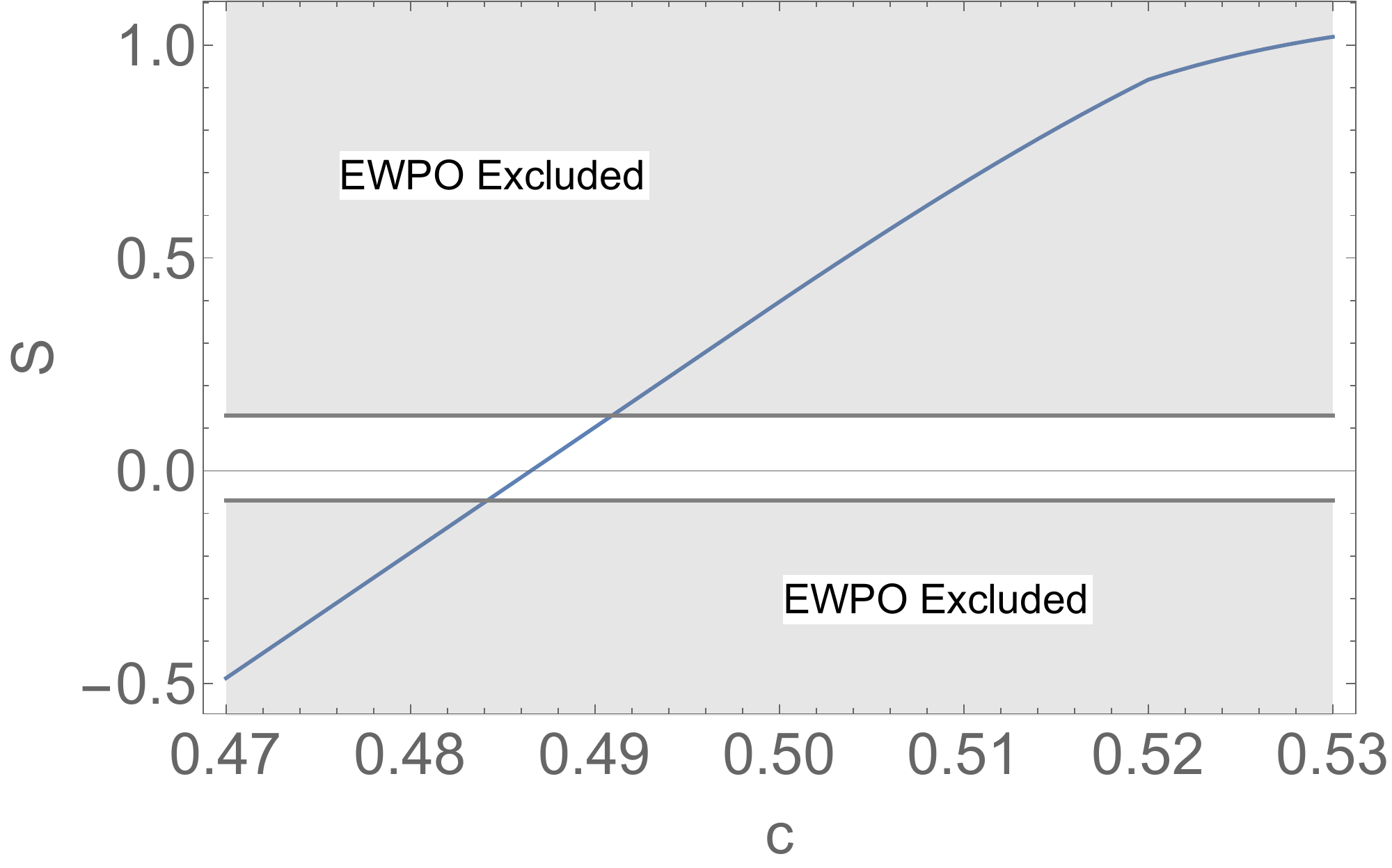}%
\includegraphics[width=0.6\textwidth]{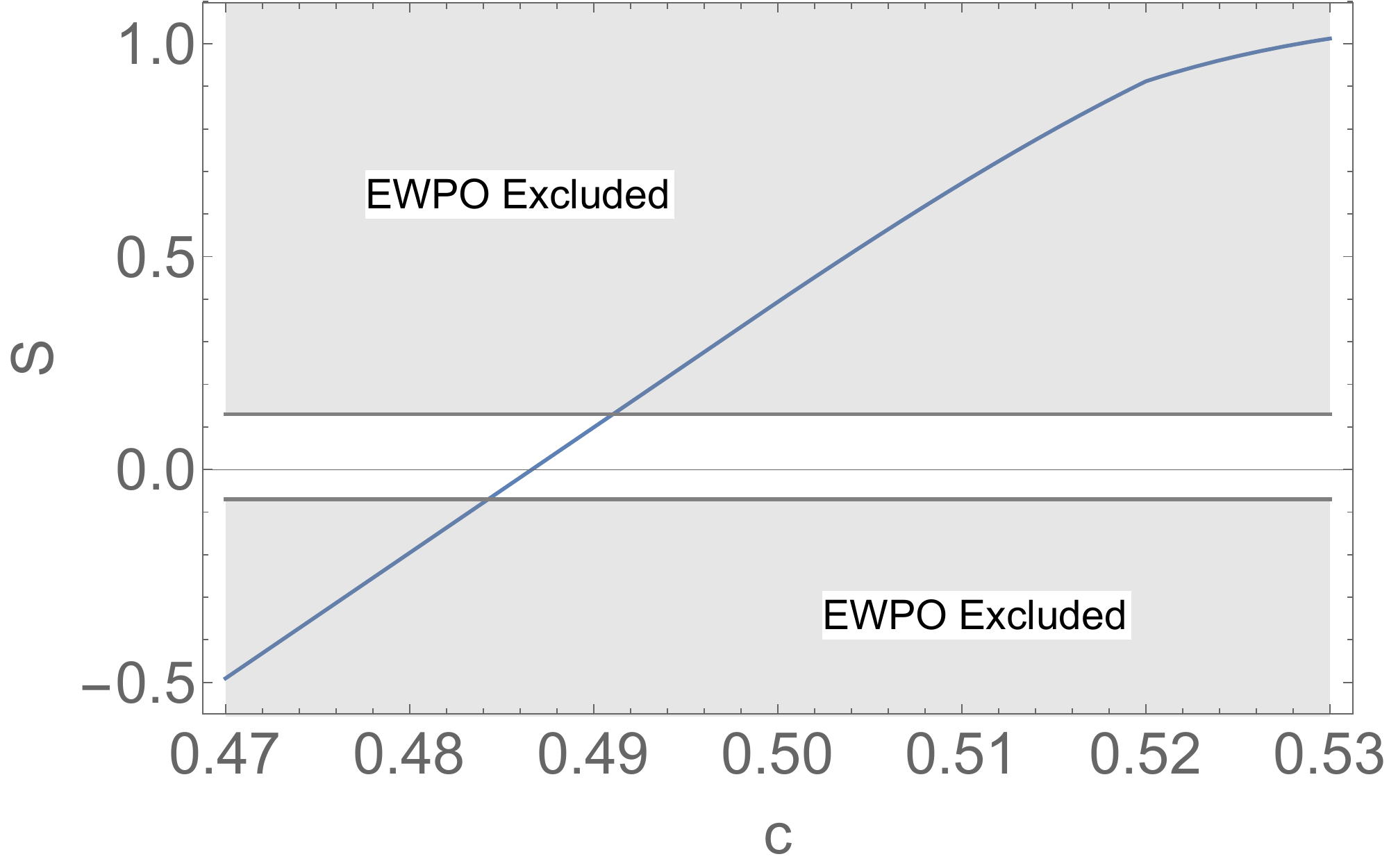}%
}%

\bigskip
\caption{The region of the fermion localization parameter $c$ that is excluded by the constraints on the $S$ parameter for $\beta = 2.1$, $R' = 1/767$ GeV$^{-1}$ (left) and $\beta = 2.5$, $R' = 1/550$ GeV$^{-1}$ }
\label{fig:EWPT}
\end{figure}

As can be seen from eq. (\ref{SCont2}), the contribution to the $S$ parameter can be made to vanish at $c \approx (1/2)(1-(\log(R'/R))^{-1}) \approx 0.487$. In Fig.~\ref{fig:EWPT}, we show the allowed region of the fermion localization parameter $c$ for our two benchmark points. The plot shows the constraints from EWPO can be avoided, but they limit the allowed values of $c$ to be in the vicinity of $\sim 0.487$.

Next, we turn our attention to the constraints from KK graviton searches. The graviton field can be obtained by expanding the fluctuations of the metric around the Minkowski space \cite{Randall:1999bb}:

\begin{equation}\label{MetricFluc}
ds^{2} = \Big(\frac{R}{z} \Big)^{2} \Big( \big( \eta_{\mu\nu}+h_{\mu\nu}(x,z) \big) dx^{\mu}dx^{\nu} -dz^{2} \Big)
\end{equation}
where $h_{\mu\nu} = \hat{h}_{\mu\nu}(x)\Psi(z)$ is the gaviton field. (\ref{MetricFluc}) can be used in Einstein's equation to find the graviton's E.O.M:

\begin{equation}\label{GravEOM}
\partial^{2}_{z} \Psi + \big( m^{2} - \frac{15}{4 z^{2}}\big) \Psi =0
\end{equation}

\noindent with Neumann boundary conditions at both branes. The solution of this equation is given by:

\begin{equation} \label{GravSol}
\Psi(z) = \sqrt{z} \big( A J_{2}(m z) + B Y_{2} (m z)  \big)
\end{equation}

We can apply the boundary conditions to find the first excited KK graviton. This givens $\sim 2.5$ TeV and $\sim 1.8$ TeV for benchmark point 1 and 2 respectively. 
The latest KK graviton searches from the LHC \cite{ATLAS6} shows that a graviton mass below 5 TeV for $k/\overline{M_{p}}$ between 0.01 and 0.3, where $k = 1/R$ is excluded. This would push the limit on $R'$ to be $\lesssim 1/1.5$ TeV$^{-1}$. Fortunately, these bounds can easily be avoided by reducing  $k/\overline{M_{p}}$. In our calculation, we used  $k/\overline{M_{p}} = 1$. If we allow $R$ to be roughly two orders of magnitude less than the Planck scale, we obtain $k/\overline{M_{p}} \lesssim 0.01$, which would suppress the couplings of the first excited KK graviton, making it possible to avoid detection the LHC searches. This does not affect our model, as it is not sensitive to the exact value of $R$.

\subsection{Radion Associated Production}

\begin{figure}[H]
\includegraphics[scale = 0.4]{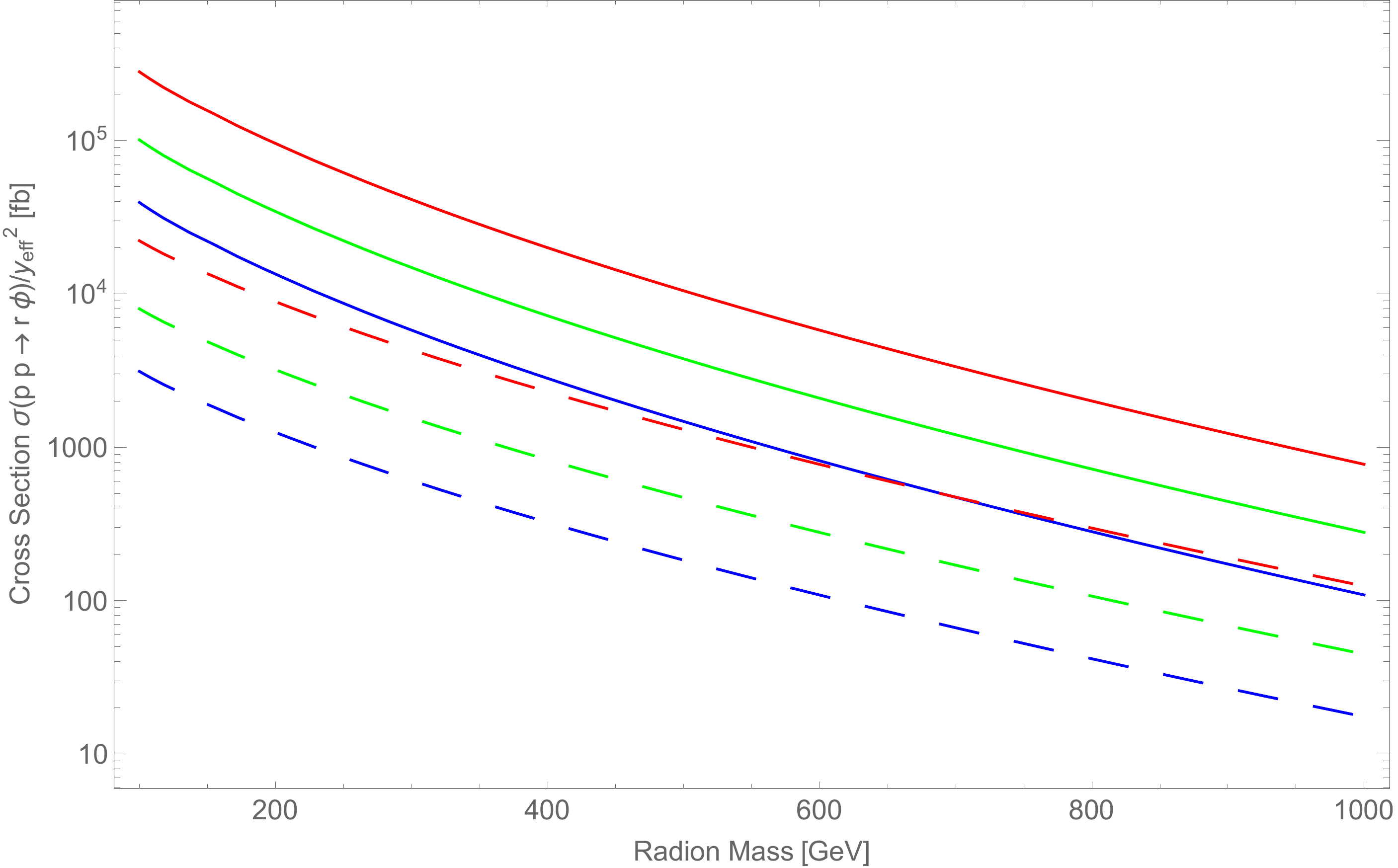}
\centering
\caption{Scalar-Radion associated production cross section with paramaters $R' = 1/500$ GeV$^{-1}$, $\beta = 2.1$, $c = 0.487$, as a function of the radion mass. We have chosen two values for the scalar mass $m_\phi$=$600$~GeV (solid) and $1000$~GeV (dashed). The three colors red, green and blue correspond to $\Lambda_{r} = 3, 5$ and $8$ TeV, respectively. }
\label{fig:Xsection}
\end{figure}

Radions can also be produced in association with a scalar either through a box diagram or through an off-shell scalar decaying to a radion and an on-shell scalar as in Fig.~\ref{fig:associated_production} below. These two processes can be used to search for the radion at the LHC.\\ 
\begin{figure}
\includegraphics[scale = 0.5]{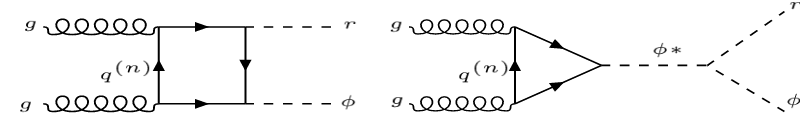}
\centering
\caption{Scalar-Radion associated production channels }
\label{fig:associated_production}
\end{figure}
The partonic cross section is given by \cite{Plehn:1996wb,Slawinska:2014vpa}
\begin{equation}
\frac{d \hat{\sigma}(gg\to\phi\,r)}{d\hat{t}}=\int d\hat{t} \frac{G_F^2\alpha_s^2}{256(2\pi)^3}\sum_{q^{(n)}} \Big(|C_\triangle F_\triangle+C_\Square F_\Square|^2+|C_\Square G_\Square|^2\Big)
\end{equation}
where $\hat{t}$ is the Mandelstam variable. In the following analysis, we consider only the case when radion mass $m_r \lesssim 1$~TeV. Thus, we will use the large fermion mass limit $4 m_f^2 \gg m_r^2, m_\phi^2$ where the form factors can be approximated as
\begin{equation}
F_\triangle=2/3,\qquad F_\Square=-2/3,\qquad G_\Square=0
\end{equation}
Meanwhile, the generalized couplings take the form
\begin{equation}
C_\triangle= \frac{2 y_{r\phi\phi}v}{\hat{s}-m_\phi^2}\frac{y_{\phi\bar{f}f} v}{m_f},\qquad C_\Square=\frac{y_{r\bar{f}f} v}{m_f}\frac{y_{\phi\bar{f}f} v}{m_f}
\end{equation}
The cross sections at different radion mass $m_r$ are presented in Fig.~\ref{fig:Xsection} for $\Lambda_r=3, 5,$ and $8$~TeV. Generally, QCD radiative corrections are particularly important. But as a rough estimation, it has been neglected in this work. The strong coupling constant $\alpha_s$ is evaluated at $\mu=m_Z$. We use MSTW 2008 \cite{Martin:2009iq} as the parton distribution functions in this calculation. 

As shown in the figure, the cross section can be significant for radion masses $\lesssim 300$ GeV. One would reasonably expect the cross section to be small due to the loops in both processes and due to the off-shell condition in the second process. In this model it is the sizable couplings that enhance the cross sections enough that their discovery are within the reach of the LHC Run II.

\section{Radion Discovery Prospects}

We now discuss the prospects of discovering the radion at the LHC using the scalar as a probe. First, we need to compare our results with the latest constraints on the radion parameter space. We find the latest constraints in \cite{Frank}, where in Fig.~$2$ in their paper they show the constraints on the $\Lambda_{r} - m_{r}$ parameter space. Their plot shows a heavily constrained region between $150$ GeV $\lesssim m_{r} \lesssim 500$ GeV, where the region of $\Lambda_{r} \lesssim 5$ TeV is mostly excluded. For lighter radion masses $\lesssim 100$ GeV, collider constraints become less relevant and astophysical bounds begin to dominate. Putting this together, the possible regions for radion searches are for $100$ GeV $\lesssim m_{r} \lesssim 150$ GeV for $\Lambda_{r} \gtrsim 1.2$ TeV, and for $m_{r} \gtrsim 150$ GeV with $\Lambda_{r} \gtrsim 5$ TeV. In either case, these bounds hardly constrain our model, since even for  $\Lambda_{r} \gtrsim 8$ TeV, the associated production cross section of a radion and a scalar remain significant. 

Fig.~$3$ in \cite{Frank} shows the decay channels and branching ratios of the radion. For radion masses $\lesssim 170$ GeV, $gg$ and $b\bar{b}$ decays are dominant, while for heavier radion masses $WW$ and $ZZ$ decays dominate. $t\bar{t}$ is triggered for $m_{r} > 2m_{t}$ and quickly rises becoming the second most dominant decay channel. 

In the range where $r\rightarrow WW$ dominate ($m_r>170$ GeV), the $\phi_0\rightarrow rr$ pair production process could be seen in searches for $4l$+missing energy. The $W\rightarrow l \nu$ branching ratio dominates, and multilepton channels have a clean signature with low SM background. For associated pair production $gg\rightarrow\phi_0 r$ with a single radion, the $2l$ channel is more promising. 

On the other hand, if the radion mass is less than $\sim 170$ GeV, QCD background becomes significant and both gluons and b quarks appear as jets. The radion pair production process in this scenario is not very promising as the signature would be $4j$. The radion associated production process, however, provides a handle for distinguishing the radion from the QCD background. If the radion is produced in association with the bulk scalar, we can use the diphoton (or $WW$ and $ZZ$) signature from the scalar as a way to trigger on $r\rightarrow b\bar{b}$ events. This signal would be detectable in searches for $jj+\gamma\gamma$ or $jj+ll +$ missing energy with a requirement for b-tagged jets.

The prime region for the radion discovery would be for a radion mass between $\sim 170$ GeV and $\sim 370$ GeV, as both processes would contribute, and the QCD background would be relatively small. In this region, the radion pair production cross section is $100 - 580$ fb (Fig.~\ref{fig:rr_production}), while the associated production cross section of the radion and the scalar is larger than 1 pb (Fig.~\ref{fig:Xsection}). Therefore these radion production processes are well within the reach of the LHC Run II.

\section{Conclusion} 
We presented a simplified RS model with a scalar singlet that only couples to KK fermions, and showed this scalar can lead to interesting LHC phenomenology, including unique signatures that could present evidence of the existence of extra dimensions. We also showed that our model is not excluded by the LHC Run I or by electroweak precision tests. Furthermore, we proposed the new scalar as a probe for discovering the elusive radion. The scalar could decay to a pair of radions, or could be produced in association with a radion. We found that both production cross sections can be significant over a wide range of parameter space and are within the reach of the LHC. In particular, a radion mass between $\sim 170$ GeV and $\sim 370$ GeV provides the optimal range for radion searches, even if $\Lambda_{r}$ is large.

\section*{Acknowledgment}
We thank John Terning for his advice and support, we also thank John Conway, John Gunion, Sekhar Chivukula, Markus Luty, Hsin-Chia Cheng and Robin Erbacher for their valuable assistance and for answering our questions. We also thank Ali Shayegan and William Kelly for their insigntful discussions.
\newpage
\appendix

\section{Estimation of the Number of KK modes} \label{NKK}

The metric background in a non-conformally flat coordinates is given by:

\begin{equation}
ds^{2} = e^{-2ky} \eta_{\mu\nu}dx^{\mu}dx^{\nu} -dy^{2}
\end{equation}
where $k$ is a scale factor $= 1/R \sim O(M_{Pl})$. The KK masses are given by:  $m_{KK} \sim k e^{kL}$, where $L$ is the size of the extra dimension. For some cutoff scale $\Lambda$, NDA yields:

\begin{equation}
N_{KK} = \frac{\Lambda}{k} = \frac{l_{5}}{\pi kL N_{c}}
\end{equation}
where $l_{5}$ is the 5D loop factor $=1/24\pi^{3}$ and $N_{c}$ is a color factor. Since the value of $kL$ that need to yield the hierarchy between the Planck scale and the EW scale is $\sim 37$, we find:

\begin{equation}
N_{KK} \simeq \frac{1}{37} \times \frac{24 \pi^{3}}{3\pi} \simeq 2
\end{equation}

\section{Definition of the Effective Couplings} \label{effective_couplings}

We mainly use the notation in \cite{Altmannshofer:2015xfo}. The Lagrangian with the explicit effective couplings $\lambda_{X}$ can be written as:

\begin{dmath}
\mathcal{L} \supset \lambda_{g} \frac{\alpha_{s}}{12 \pi v} \phi_{0} G_{\mu\nu}^{a}G^{a \mu \nu} + \lambda_{\gamma} \frac{\alpha}{\pi v} \phi_{0} F_{\mu \nu} F^{\mu \nu} + \lambda_{Z} \frac{\alpha}{\pi v} \phi_{0} Z_{\mu \nu} Z^{\mu \nu} + \lambda_{Z \gamma} \frac{\alpha}{\pi v} \phi_{0} Z_{\mu \nu}F^{\mu \nu} + \lambda_{W} \frac{2 \alpha}{\pi s_{w}^{2} v} \phi_{0} W_{\mu\nu}^{+} W^{-\mu\nu}
\end{dmath}

\begin{equation}
\lambda_{\gamma} = \lambda_{B} + \lambda_{W}
\end{equation}

\begin{equation}
\lambda_{W} =y_{\text{eff}} \sum_{n} \frac{1}{6}  \frac{v}{m_{f}} C_{w}(r_{f}) D_{c}(r_{f}) A_{f}(\tau_{f})
\end{equation}

\begin{equation}
\lambda_{B} =y_{\text{eff}} \sum_{n} \frac{1}{6}  \frac{v}{m_{f}} Y_{f}^{2} D_{w}(r_{f}) D_{c}(r_{f}) A_{f}(\tau_{f})
\end{equation}

\begin{equation}
\lambda_{g} =y_{\text{eff}} \sum_{n} 2  \frac{v}{m_{f}} C_{c}(r_{f}) D_{w}(r_{f})  A_{f}(\tau_{f})
\end{equation}

\begin{equation}
\lambda_{Z} = \lambda_{W} \frac{c_{w}^{2}}{s_{w}^{2}} + \lambda_{B} \frac{s_{w}^{2}}{c_{w}^{2}}
\end{equation}

\begin{equation}
\lambda_{Z\gamma} = 2 \Big( \lambda_{W} \frac{c_{w}}{s_{w}} - \lambda_{B} \frac{s_{w}}{c_{w}} \Big)
\end{equation}
\\
where the sum goes over the KK modes of all fermions in the loop. Here $Y_{f}$ is the hypercharge, $m_{f}$ is the mass of the KK fermion, $s_{w}$ and $c_{w}$ are the sine and cosine of the weak angle, $C_{w}(r_{f})$ is the index of the $SU(2)_{L}$ representation, $Tr(T^{i}T^{j}) = C_{w} \delta^{ij}$, $C_{w} = I_{f}(I_{f}+1)D_{w}(r_{f})/3$ for $D_{w} = 2I_{f}+1$ dimensional representation of $SU(2)_{L}$ and $Tr(T^{a}T^{b}) = C_{c}(r_{f})\delta^{ab}$.\\

The function $A_{f}(\tau)$ is given by:
\begin{equation}
A_{f}(\tau) = \frac{3}{2\tau^{2}} \Big[(\tau -1)f(\tau)+\tau\Big]
\end{equation}
$$
f(x) = \left\{
        \begin{array}{ll}
            \Big[ sin^{-1}(\sqrt{\tau})\Big]^{2} & \quad \tau \leq 1 \\ 
            \\
            -\frac{1}{4} \Bigg[ log \Big( \frac{1+\sqrt{1-\tau^{-1}}}{1-\sqrt{1-\tau^{-1}}} -i\pi \Big) \Bigg]^{2} & \quad \tau > 1
        \end{array}
    \right.
$$

where $\tau_{i} = \frac{m_{\phi}^{2}}{4m_{i}^{2}}$.

\setcounter{equation}{0}
\setcounter{footnote}{0}

\FloatBarrier

\end{document}